\documentclass[12pt]{article}
\newtheorem{Def}{Definition}
\newtheorem{theorem}{Theorem}

\def\sqr#1#2{{\vcenter{\vbox{\hrule height.#2pt
               \hbox{\vrule width.#2pt height#1pt \kern#1pt
                      \vrule width.#2pt}
                      \hrule height.#2pt}}}}

\def\qed{\hbox{\hskip 6pt\vrule width6pt height7pt depth1pt \hskip1pt}}
\def\R{\real}
\def\C{\complex}
\def\nd{\noindent}
\def\real{{\rm I\kern-.2em R}}
\def\complex{\kern.1em{\raise.47ex\hbox{ $\scriptscriptstyle
|$}}\kern-.40em{\rm
C}}
\def\be{\begin{equation}}
\def\ee{\end{equation}}
\def\bearray{\begin{eqnarray}}
\def\eearray{\end{eqnarray}}
\def\bra{\langle}
\def\ket{\rangle}

\title{Generic Jumps of Fredholm Indices and the Quantum Hall Effect}
\author{J.~E.~Avron and L.~Sadun\footnote{On leave from the Department of Mathematics,
University of Texas, Austin, TX 78712 USA}
\\ Department of Physics, Technion, 32000 Haifa, Israel}
\begin{document}
\maketitle
\begin{abstract}
We describe  the generic behavior of Fredholm indices in the space
of Toeplitz operators. We relate this behavior to certain
conjectures and open problems that arise in the context of the
Quantum Hall Effect.

\end{abstract}

\input epsf

\section{Introduction and Motivation}
Suppose one interpolates between  Fredholm operators with
different indices. What can one say about the way the indices
change? The answer to this question depends on the choice of the
embedding space for the Fredholm operators in question. In the
space of bounded operators, little can be said. But, in the space
of Toeplitz operators, (and then also for Toeplitz modulu
compacts), as we shall explain, the indices change by abrupt
discontinuous jumps that tend to be small. We relate this behavior
to certain conjectures and open problems that arise in the context
of the Quantum Hall Effect (QHE) \cite{qhe}.

\subsection{Physical background}

In the theory of the integer quantum Hall effect (of
non-interacting electrons) \cite{bel,ass} one identifies the Hall
conductance with the Fredholm index of a rather special operator,
namely $PUP$, thought of as an operator on the range of $P$.
Here $P=P(E)$ is an (infinite dimensional) projection in the Hilbert space
$L^2(C)$, namely the projection on the spectrum of
the one electron Hamiltonian below the Fermi energy $E$. $U$ is
the multiplication operator $z\over |z|$ associated with a
singular gauge transformation that introduces an Aharonov-Bohm
flux tube at the origin of the Euclidean plane. $PUP$ is  Fredholm
provided the integral kernel of the projection, $p(z,z';E)$ has
good decay properties as $|z-z'|$ gets large \cite{ass}.

Recent progress in the rigorous theory of random Schr\"odinger
operators relevant to the QHE \cite{aizenman} guarantees good
decay properties for $p(z,z';E)$ provided $E$ lies in certain
energy intervals. Percolation arguments \cite{trugman} and scaling
theories of localization \cite{khmel} give theoretical evidence
that these decay properties persist for all but a discrete set of
energies. This implies that the
graph of the Hall conductace as a function of $E$ should be a step
function. Indeed, experimentally, the Hall conductance in the
integer Hall effect, is close to a monotonic step function with
$\pm 1$ and  $\pm 2$ jumps \cite{laugh}. (Jumps by 2  occur when
the Hall conductance is larger than 6 and is attributed to the
smallness of the magnetic moment of the electron in these systems.)

 The smallness of the jumps of the Fredholm indices
in the QHE might, of course, be a special property of a special
system. Here, instead, we want to explore the opposite point of
view, namely the possibility that the existence of steps and the
smallness of the jumps reflects a generic property  of Fredholm
indices and has little to do with the specific properties of the
system in question.

Some support to this point of view comes from the relation of
Chern numbers and Fredholm indices. In non-commutative geometry
\cite{connes} Chern numbers and Fredholm indices are intimately
related. This is also the case in the index theory of elliptic
operators \cite{atiyah}. For Chern numbers that arise from studies
of spectral bundles (of Hamiltonians with discrete spectra), a
generic deformation of the Hamiltonian leads to a step function
with $\pm 1$ jumps in the first Chern number \cite{simon}. This
follows from the Wigner von Neumann codimension 3 rule for
eigenvalue crossing \cite{wvn} and the fact that a generic
crossing is a conic crossing and is not system specific.

As far as the QHE goes one might argue that since the Hall
conductance can be directly related to a Chern number
\cite{qhe,tknn}, the genericity of small jumps follows immediately.
The difficulty with this argument has to do with the thermodynamic limit.
Normally, the QHE is associated with large systems. The genericity
result quoted above for Chern numbers is for operators with
discrete spectrum. This is the case for finite sytems, but is in
general not the case for extended systems, and in particular does not apply to
models of the quantum Hall effect. The main attractive feature of
the Fredholm approach to the Hall effect is that it is phrased
directly in the thermodynamic limit.

 Another way of
phrasing the main theme of this paper is: What, if any, is the
analog for Fredholm operators of the genericity of small jumps in
Chern numbers?

\subsection{The mathematical problem}

We wish to interpolate between two (or more) Fredholm
operators.  If the indices of these operators are different this
cannot be done within the space of Fredholm operators. At some
points in the interpolation the Fredholm property will be lost and
the index will be ill defined. For ``generic'' interpolations,
what is the nature of this bad set?  Near such a bad point,
how big a range of indices can be found?

Working in the space of bounded operators, little can be said. The
space is simply too large, and when the Fredholm property is lost
we lose all analytic control. However, in the  space of
sufficiently smooth Toeplitz operators interesting results can be
obtained. In systems without symmetry, we find the following
behavior: Almost every operator is Fredholm, and sets of
codimension $n$ appear as boundaries between regions of Fredholm
operators whose indices differ by $n$.  We speak simply of the
index ``jumping by $n$'' on a set of codimension $n$.

In systems with a $Z_2$ symmetry (e.g. time reversal symmetry or
complex conjugation symmetry), sets of codimension $n$ appear as
common boundaries of regions of Fredholm operators whose indices
differ by as much as $2n$.  That is, the index can jump by as much
as $2n$ on a set of codimension $n$.

\section{Basic Definitions and Properties}

We review here the basic definitions and properties of Fredholm
operators on separable Hilbert spaces. For a more complete
treatment see \cite{douglas}.

\begin{Def} A bounded operator $A$ on a separable Hilbert space if {\em Fredholm} if there
exists another bounded operator $B$ such that $1-AB$ and $1-BA$ are compact.
\end{Def}

In particular, the kernel and cokernel of $A$ are finite dimensional, and we define

\begin{Def} The {\em index} of a Fredholm operator $F$ is
\be Index(F)= dim\, Ker(F)- dim\,Ker(F^\dagger).\ee
\end{Def}

Fredholm operators are stable under compact perturbations and under small bounded perturbations.
That is, if $A$ is Fredholm, there exists an $\epsilon >0$ such that, for any bounded operator
$B$ with operator norm $\| B \| < \epsilon$ and for any compact operator $K$, the operator
$A+B+K$ is Fredholm with the same index as $A$.

The simplest example of a Fredholm operator with nonzero index is the shift operator.
Let $e_0, e_1, e_2, \ldots$ be an orthonormal basis for a Hilbert space, and let the operator
$a$ act by
\be
a(e_n) = \cases{e_{n-1} & if $n>0$ \cr 0 & if $n=0$}.
\ee
The adjoint of $a$ acts by
\be a^\dagger(e_n) = e_{n+1}
\ee
Since $a a^\dagger= a^\dagger a + |e_0 \ket \bra e_0 |$ is
 the identity, $a$ is Fredholm.
The kernel of $a$ is 1-dimensional. The cokernel of $a$,
which is the same as the kernel of
$a^\dagger$, is 0 dimensional.  Thus the index of $a$ is 1.
 Similarly, $a^\dagger$ is Fredholm
with index $-1$.

The following theorem is standard:
\begin{theorem} If $A_1, \ldots A_n$ are Fredholm operators, then
the product $A_1 A_2 \cdots A_n$ is also
Fredholm, and $Index(A_1\cdots A_n) = \sum_{i=1}^n Index(A_i)$.
\end{theorem}

Finally we consider connectedness in the space of Fredholm
operators. If $A$ and $A'$ are Fredholm operators on the same
Hilbert space, then there is a continuous path of Fredholm
operators from $A$ to $A'$ if and only if $Index(A)=Index(A')$.
(By continuous, we mean relative to the operator norm).  Put
another way, the path components of $Fred(H)$, the space of
Fredholm operators on $H$, is indexed (pun intended) by the
integers.  The $n$-th path component is precisely the set of
Fredholm operators of index $n$ \cite{douglas}.

\section{Fredholm Operators in the Space of Bounded Operators}

The most natural setting for our problem is consider arbitrary bounded operators, with the
topology defined by the operator norm.  We ask how many parameters must be varied in order
to reach the common boundary of two regions, whose indices differ by $k$.  Unfortunately,
the answer is independent of $k$:

\begin{theorem} Let $U_n$ be the set of Fredholm operators of index $n$. Every point on the
boundary of $U_n$ is also on the boundary of $U_m$, for every integer $m$.
\end{theorem}

Proof: Let $A$ be a (not Fredholm) operator on the boundary of $U_n$.  Given $\epsilon >0$,
we must find an operator in $U_m$ within a distance $\epsilon$ of $A$.

Suppose that the kernel and cokernel of $A$ are infinite
dimensional, and that there is a gap in the spectrum of $A^\dagger
A$ at zero.  (If this is not the case, we may perturb $A$ by an
arbitrarily small amount to make it so).  Now let $B$ be a unitary
map from the kernel of $A$ to the cokernel.   Let $P,\ (P')$ be
the orthogonal projection onto $ker(A), \ (coker(A))$, and let $a$
be a shift operator on $ker(A)$. For each $m \ge 0$,
$A(\epsilon)=A + \epsilon B a^m P$ has a bounded right inverse
\be
A^\dagger {1\over P'+ AA^\dagger} P'_\perp + {1\over\epsilon}
(a^\dagger)^m B^\dagger P'.
\ee
It follows that the cokernel of
$A(\epsilon)$ is empty. It is easy to see that the kernel of
$A(\epsilon)$ is $m$ dimensional hence $Index(A(\epsilon))=m$.
Similarly, $A + \epsilon B (a^\dagger)^{m} P$ has index $-m$.
\hfill$\qed$

This theorem tells us that, in the space of all bounded operators
 there is no specific notion of
being at a transition point from index $n$ to index $m$. As long
as  an operator stays Fredholm, its index cannot change, and when
it fails to be Fredholm it can change into anything.

To achieve useful results, we must work on a smaller space.

\section{Linear Combinations of  Shifts}

In this section and the next we show that ``generic" behavior is
indeed achieved in some finite dimensional spaces, and in some
infinite-dimensional spaces with sufficiently fine topologies.  We
see also how control is lost as the space is enlarged and the
topology is coarsened.

\subsection{Shift by one}

We begin by considering linear combinations of the shift operator $a$ and the identity
operator 1.  That is, we consider the operator
$$ A = c_1 a + c_0 $$
where $c_1$ and $c_0$ are constants.

\begin{theorem}
\label{0-1}
If $|c_1| \ne |c_0|$, then $A$ is Fredholm.
The index of $A$ is 1 if $|c_1|>|c_0|$ and zero if $|c_1|<|c_0|$.
If $|c_1| = |c_0|$, then $A$ is not Fredholm.
\end{theorem}

\nd Proof: First suppose $|c_0| > |c_1|$.  Then $A$ is invertible:
$$ A^{-1} = c_0^{-1} (1 + (c_1/c_0) a)^{-1} = \sum_{n=0}^\infty {(-1)^n c_1^n
\over c_0^{n+1}} a^n,
$$
as the sum converges absolutely.  Thus $A$ has neither kernel nor cokernel, and has
index zero.

If $|c_1| > |c_0|$, then the kernel of $A$ is 1-dimensional, namely all multiples of
$|\psi \ket = \sum_{n=0}^\infty z_0^n e_n$, where $z_0 = -c_0/c_1$. Notice how the norm
of $|\psi \ket$ goes to infinity as $|z_0| \to 1$.  However, $A^\dagger$ has no kernel,
since for any unit vector $|\phi \ket$,
$\|A^\dagger | \phi \ket\| = \|\bar c_1 a^\dagger |\phi \ket + \bar c_0 |\phi \ket\|
\ge   \|\bar c_1 a^\dagger |\phi \ket \| - \|\bar c_0 |\phi \ket\|
= |c_1| - |c_0|$.  Thus the index of $A$ is 1.

If $|c_1| = |c_0| $, then $A$ is at the boundary between index 1
and index 0, and so cannot be Fredholm.\hfill$\qed$

\subsection{Finite linear combinations of shifts}

Next we consider linear combinations of $1, a, a^2, \ldots$ up to some fixed $a^n$.
That is, we consider operators of the form
\be A = c_n a^n + c_{n-1} a^{n-1} + \cdots + c_0.
\label{A}
\ee
This is closely related to the polynomial
\be p(z) = c_n z^n + \cdots + c_0.
\label{p}
\ee

\begin{theorem}
If none of the roots of $p$ lie on the unit circle, then $A$ is Fredholm, and the
index of $A$ equals the number of roots of $p$ {\em inside} the unit circle, counted with
multiplicity.  If any of
the roots of $p$ lie {\em on} the unit circle, then $A$ is not Fredholm.
\end{theorem}

\nd Proof: The polynomial $p(z)$ factorizes as $p(z) = c_k \prod_{i=1}^k (z-\zeta_i)$, where $k$
is the degree of $p$ (typically $k=n$, but it may happen that $c_n=0$).  But then
$A = c_k \prod_{i=1}^k (a - \zeta_i)$. If none of the roots $\zeta_i$ lie on the unit circle,
then each term in the product is Fredholm, so the product is Fredholm, and the index of
the product is the sum of the indices of the factors.  By Theorem \ref{0-1}, this exactly
equals the number of roots $\zeta_i$ inside the unit circle.

If any of the roots lie on the unit circle, then a small
perturbation can push those roots in or out, yielding Fredholm
operators with different indices. This borderline operator
therefore cannot be Fredholm.\hfill$\qed$

The last theorem easily generalizes to linear combination of left-shifts and right-shifts.
The index of an operator
\be A = c_n a^n + \cdots + c_1 a + c_0 + c_{-1} a^\dagger + \cdots + c_{-m} (a^{\dagger})^m
\ee
equals the number of roots of
\be p(z) = \sum_{i=-m}^n c_i z^i
\ee
inside the unit circle, minus the degree of the pole at $z=0$ (that is $m$, unless $c_{-m}=0$).
This follow from the fact that
\be A = (\sum_{i=-m}^n c_i a^{i+m}) (a^\dagger)^m.
\ee

Since there is no qualitative difference between combinations of left-shifts and combinations
of both left- and right-shifts, we restrict our attention to left-shifts only, and consider
families of operators of the form (\ref{A}).

\begin{theorem}\label{finite-combinations}
In the space of complex linear combinations of 1, $a$, \ldots, $a^n$, almost every operator is
Fredholm.  For every $k \le n$, the points where the index can jump by $k$ (by which we mean
the common boundaries of regions of Fredholm operators whose indices differ by $k$) is a set
of real codimension $k$.

In the space of real linear combinations of 1, $a$, \ldots, $a^n$, almost every operator is
Fredholm.  For every $k \le n$, the points where the index jumps by $k$ is a stratified space,
the largest stratum of which has
 real codimension $\lfloor (k+1)/2 \rfloor $, where $ \lfloor x \rfloor$ denotes the integer
part of $x$.
\end{theorem}

\nd Proof:
Our parameter space is the space of coefficients $c_i$, or equivalently the space of
polynomials of degree $\le n$. This is either $\R^{n+1}$ or $\C^{n+1}$, depending on whether
we allow real or complex coefficients.
In either case, the set $U_k$ of Fredholm operators of index $k$ is identical to the set of
polynomials with $k$ roots inside the unit circle and the remaining $n-k$ roots outside
(if $c_n=0$, we say there is a root at infinity; if $c_n=c_{n-1}=0$, there is a double root
at infinity, and so on. Counting these roots at infinity, there are always exactly $n$ roots
in all.) The boundary of
$U_k$ is the set of polynomials with at most $k$ roots inside the unit circle, at most $n-k$
outside the unit circle, and at least one root on the unit circle. (Strictly speaking, the
zero polynomial is also on this boundary.  This is of such high codimension that it has no
effect on the phase portrait we are developing.).   We consider the common
boundary of $U_k$ and $U_{k'}$.  If $k<k'$, a nonvanishing polynomial is on the boundary of
both $U_k$ and $U_{k'}$ if it has at most $k$ roots inside the unit circle and at most
$n-k'$ roots outside.  It must therefore have at least $k'-k$ roots on the unit circle.

If we are working with complex coefficients, this is a set of codimension $k'-k$. The roots
themselves, together with an overall scale $c_n$, can be used to parametrize the space of
polynomials.  For each root, being on the unit circle is codimension 1, while being inside or
outside are open conditions.  Since the roots
are independent, placing $k'-k$ roots on the unit circle is codimension $k'-k$.

If we are working with real coefficients, the roots are not independent, as non-real roots come
in complex conjugate pairs.  Thus, the common boundary of $U_k$ and $U_{k'}$ breaks into several
strata, depending on how many real roots and how many complex conjugate pairs lie on the unit
circle.  If $k'-k$ is even, the biggest stratum consists of having $(k'-k)/2$ pairs, and has
codimension $(k'-k)/2$.  If $k'-k$ is odd, the biggest stratum consists of having $(k'-k-1)/2$
pairs and one real root on the unit circle, and has codimension $(k'+1-k)/2$.  \hfill\qed

\begin{figure}
\vskip -1in
\centerline{\hskip -1.3in\epsfxsize=3.0truein \epsfbox{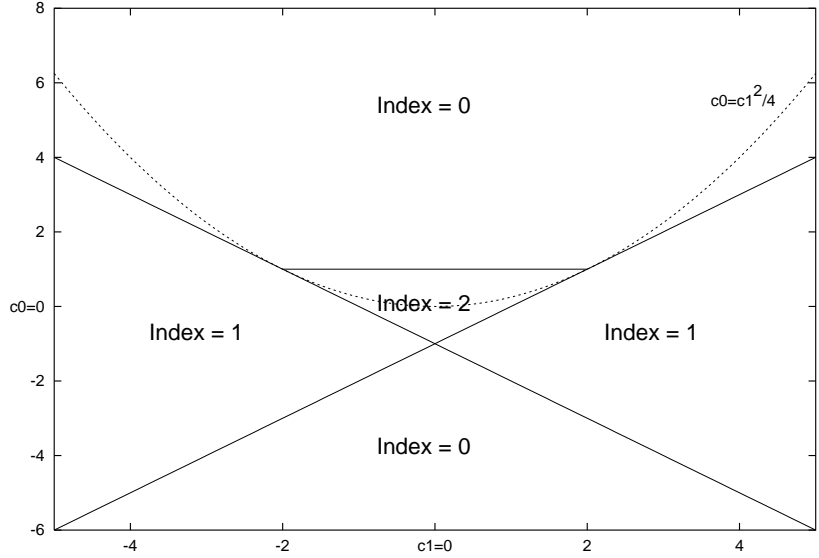}}
\caption{A phase plot for $A=a^2 + c_1 a + c_0$.}
\label{fig1}
\end{figure}
%\centerline{\bf Figure 1. A phase plot for $A=a^2 + c_1 a + c_0$.}

Theorem \ref{finite-combinations} is illustrated in Figure~1,
where the phase portrait is shown for $n=2$ with real
coefficients, with $c_2$ fixed to equal 1. The points above the
parabola $c_0 = c_1^2/4$ have complex conjugate roots, while
points below have real roots. Notice that the transition from
index 2 to index 0 occurs at an isolated point when the roots are
real, but on an interval when the roots come in complex-conjugate
pairs.

It is clear that an almost identical theorem applies to linear combinations of left-shifts up
to $a^n$ and right-shifts up to $(a^\dagger)^m$.  The results are essentially independent
of $n$ and $m$ (their only effect being to limit the size of possible jumps to $n+m$). We can
therefore extend the results to the space of all (finite) linear combinations of left- and
right-shifts, which is topologized as the union over all $n$ and $m$ of the spaces considered
above. Our result, restated for that space, is

\begin{theorem}\label{limit-space}
In the space of finite complex linear combinations of left- and right-shifts of arbitrary degree,
almost every operator is Fredholm.  For every integer $k \ge 1$, the points where the index can
jump by $k$ (by which we mean
the common boundaries of regions of Fredholm operators whose indices differ by $k$) is a set
of real codimension $k$.

If we restrict the coefficients to be real, then, for every $k \le n$, the points where
the index jumps by $k$ is a stratified space,
the largest stratum of which has real codimension $\lfloor (k+1)/2 \rfloor $.
\end{theorem}

\section{Toeplitz operators}

Although Theorem \ref{limit-space} refers to an infinite-dimensional space, this space is
still extremely small -- each point is a {\em finite} linear combination of shifts. In this
section we consider {\em infinite} linear combinations of shifts. This is equivalent to
studying Toeplitz operators.

\begin{Def} The {\em Hardy space} $H$ is the subspace of $L^2(S^1)$ consisting of functions
whose Fourier transforms have no negative frequency terms.  Equivalently, if we give $L^2(S^1)$
a basis of Fourier modes $e_n = e^{in\theta}$, where the integer
$n$ ranges from $-\infty$ to $\infty$, then $H$ is the closed linear span of $e_0, e_1, e_2,
\ldots$.
\end{Def}

We think of $S^1$ as sitting in the complex plane, with $z=e^{i\theta}$. Now let $f(z)$ be a
bounded, measurable function on $S^1$, and let $P$ be the orthogonal projection from $L^2(S^1)$
to $H$.  If $| \psi \ket \in H$, then $|f \psi\ket$ (pointwise product) is in $L^2(S^1)$, and
$P|f \psi\ket \in H$.
We define the operator $T_f$ by
\be T_f |\psi\ket = P|f \psi\ket. \label{Toep}
\ee

\begin{Def} An operator of the form (\ref{Toep}) is called a {\em Toeplitz operator}. We call
a Toeplitz operator $T_f$ continuous if the underlying function
$f$ is continuous, and apply the terms ``differentiable",
``smooth" and ``analytic" similarly.
\end{Def}

\nd {\bf Remark}: Toeplitz operators can be represented by semi-infinite
matrices that have constant entries on diagonals, and the various
classes we have defined correspond to the decay away from the main
diagonal.

Notice that \be T_{e_m} e_n = \cases{ e_{n+m} & if $n+m \ge 0$ \cr
0 & otherwise} \ee so $T_{e_m}$ is simply a shift by $m$, a right
shift if $m>0$ and a left-shift if $m<0$. All our results about
shifts can therefore be understood in the context of Toeplitz
operators. Theorem \ref{finite-combinations} refers to operators
$T_f$, where $f$ is a polynomial in $z^{-1}$ of limited degree. Theorem
\ref{limit-space} considers polynomials or arbitrary degree in $z$
and $z^{-1}$.  We will see that the results carry over to analytic
functions on an annulus around $S^1$, and to a lesser extent to
$C^k$ Toeplitz operators, but with results that weaken as $k$ is
decreased.

Here are some standard results about Toeplitz operators.
 For details, see \cite{douglas}.

\begin{theorem} A $C^1$ Toeplitz operator $T_f$ is Fredholm if and only if $f$ is
everywhere nonzero on the unit circle.  In that case the index of $T_f$ is minus the winding
number of $f$ around the origin, namely
\be Index(T_f) = -Winding(f) = {-1\over 2 \pi i} \int_{S^1} {df \over f}, \label{wind}
\ee
\end{theorem}

Given the first half of the theorem, the equality of index and
winding number is easy to understand.  We simply deform $f$ to a
function of the form $f(z)=z^n$, while keeping $f$ nonzero on all
of $S^1$ throughout the deformation (this is always possible, see
e.g. \cite{gp}). In the process of deformation, neither the index
of $T_f$ nor the winding number of $f$ can change, as they are
topological invariants.  Since the winding number of $z^n$ is $n$,
and since $T_{z^n} = (a^\dagger)^n$ (if $n\ge 0$, $a^{-n}$
otherwise), which has index $-n$, the result follows.

We now consider functions $f$ on $S^1$ that can be analytically
continued (without singularities) an  annulus $r_0 \le |z| \le
r_1$, where the radii $r_0 < 1$ and $r_1>1$ are fixed. This is
equivalent to requiring that the Fourier coefficients $\hat f_n$
decay exponentially fast, i.e. that the sum
\be
\sum_{n=-\infty}^\infty |\hat f_n |(r_0^n + r_1^n)
\ee
converges. For now we do not impose
any reality constraints or other symmetries on the coefficients
$\hat f_n$.  This space of functions is a Banach space, with norm
given by the sup norm on the annulus.  This norm is stronger than
any Sobolev norm on the circle itself.

The analysis of the corresponding Toeplitz operators is straightforward and similar
to the proof of Theorem \ref{finite-combinations}.  Since $f$ has no poles
in the annulus, we just have to keep track of the zeroes of $f$.  For the index of $T_f$
to change, a zero of $f$ must cross the unit circle.  For the index to jump from $k$ to $k'$,
$|k-k'|$ zeroes must cross simultaneously.  In the absence of symmetry, the locations of the
zeroes are independent and can be freely varied, so this is a codimension-$|k-k'|$ event.

If we impose a reality condition: $f(\bar z) = \overline{f(z)}$, then zeroes appear only on
the real axis or in complex conjugate pairs. In that case, changing the index by 2 is merely
a codimension-1 event.  Combining these observations we obtain

\begin{theorem}\label{analytic}
In the space of Toeplitz operators that are analytic in a (fixed)
annulus containing $S^1$, almost every operator is Fredholm.  For
every integer $k \ge 1$, the points where the index can jump by
$k$ is a set of real codimension $k$.

If we impose a reality condition $f(\bar z)=\overline{f(z)}$ then,
for every $k \le n$, the points where
the index jumps by $k$ is a stratified space,
the largest stratum of which has real codimension $\lfloor (k+1)/2 \rfloor $.
\end{theorem}

Finally we consider Toeplitz operators that are not necessarily analytic, but are merely
$\ell$ times differentiable, and we use the $C^\ell$ norm.  Our result is

\begin{theorem}\label{C^ell}
In the space of Toeplitz $C^\ell$ operators, almost every operator is Fredholm.
For every integer $k$ with $1 \le k \le 2\ell+1$, the points where the index can
jump by $k$ is a set of real codimension $k$.  For every integer $k \ge 2\ell+1$, the points
where the index can jump by $k$ is a set of real codimension $2\ell+1$.
\end{theorem}

In other words, our familiar results hold up to codimension $2\ell+1$, at which point we lose
all control of the change in index.

\smallskip

\nd Proof: As long as $f$ is everywhere nonzero, $T_f$ is Fredholm.  To get a change in
index, therefore, we need one or more points where $f$, and possibly some derivatives of
$f$ with respect to $\theta$, vanish.  Suppose then that
for some angle $\theta_0$, $f(\theta_0)=f'(\theta_0)=\cdots=f^{(n-1)}(\theta_0)=0$ for
some $n \le \ell$, but that the $n$-th derivative $f^{(n)}(\theta_0) \ne 0$. This is
a codimension $2n-1$ event, since we are setting the real and imaginary parts of $n$
variables to zero, but have a 1-parameter choice of points where this can occur.
Without loss of generality, we suppose that this $n$-th derivative is real and positive.
By making a $C^\ell$-small perturbation of $f$, we can make the value of $f$ highly oscillatory
near $\theta_0$, thereby wrapping around the origin a number of times.  However, since a
$C^\ell$-small perturbation does not change the $n$-th derivative by much, the sign of the
real part of $f$ can change at most $n$ times near $\theta_0$, so the argument of $f$ can only
increase or decrease by $n\pi$ or less.  The difference between these two extremes is $2n\pi$,
or a change in winding number of $n$.

To change the index by an integer $m$, therefore, we must have the function vanish to various
orders at several points, with the sum of the orders of vanishing adding to $m$. The generic
event is for $f$ (but not $f'$) to vanish at $m$ different points -- this is a codimension $m$
event, analagous to having $m$ zeroes of a polynomial cross the unit circle simultaneously at
$m$ different points. All other scenarios have higher codimension and are analogous to
having 2 or more zeroes of the $m$ zeroes crossing the unit circle at the same point.

The situation is different, however, when the function $f$ and the
first $\ell$ derivatives all vanish at a point $\theta_0$. Then
the higher-order derivatives are not protected from $C^\ell$-small
perturbations and, by making such a perturbation, we can change
$f$ into a function that is identically zero on a small
neighborhood of $\theta=\theta_0$. By making a further small
perturbation, we can make $f$ wrap around the origin as many times
as we like near $\theta=\theta_0$. More specifically, if $f$ is
zero on an interval of size $\delta$, then, for small $\epsilon$,
$\tilde f(\theta) = f(\theta) + \epsilon e^{iN\theta}$ will wrap
around the origin approximately $N\delta/2\pi$ times near
$\theta_0$.  By picking $N$ as large (positive or negative) as we
wish, we can obtain arbitrarily positive or negative indices.  As
long as we take $\epsilon \ll N^{-\ell}$, this perturbation will
remain small in the $C^\ell$ norm.\hfill\qed

\section{The Quantum Hall Effect}

We have seen in the previous section that the Fredholm index of a
generic one dimensional family of Toeplitz operators is a step
function with small jumps. This is reminiscent of what one
observes for the Hall conductance for random Schr\"odinger
operators.

 In this section we want to discuss some of the
difficulties, and what one would still need to know, for the
strategy in this paper to yield useful results for the QHE.

\subsection{Landau levels}
The Hall conductance is related to the Index of $PUP$ (on $Range\
P$) with $P$ a spectral projection in $L^2(\complex)$ and $U$ a
multiplication by $z\over |z|$. This operator is closely related
to a Toeplitz operator in the case of a basic paradigm for the
Hall effect:
\begin{theorem} Let $P$ be a projection on the lowest Landau level
in $\real^2$, then $PUP$ differs from a Toeplitz operator by a
compact operator.
\end{theorem}
\nd Proof: A basis for the lowest Landau level is
\be
|n\ket={1\over \sqrt{\pi\, n!}}\, z^n\,e^{-|z|^2/2},\quad n\ge 0.
\ee As a consequence
\be
\bra n|U|m\ket=\delta_{n,m+1}\, {(m+1/2)!\over
m!\sqrt{m+1}}\approx \delta_{n,m+1}\left(1-{1\over 8m}\right).
\qed \ee

The same result also holds if $P$ is a projection on a higher Landau level,
but the calculation is more involved. If $P$ is a projection onto multiple
Landau levels, then $PUP$ is a compact perturbation of a direct sum of
Toeplitz operators, one for each Landau level.
This suggests that the class of Toeplitz operators is indeed
related to the QHE.

 For (spinless) electrons/holes on the Euclidean and hyperbolic planes,
with homogeneous magnetic field, and without disorder, $Index
(PUP)(E)$ has been explicitly computed as a function of the
``Fermi energy'' $E$. In the Euclidean plane one finds a monotonic step function with
jumps $\pm 1$ \cite{pnueli}. (One needs both signs for electrons and
holes.) The same results apply in the hyperbolic plane for all energies below the continuous
spectrum \cite{pnueli}.  This implies that also for (relatively) compact
perturbations of these Hamiltonians the Fredholm index in the QHE
behaves as does the Fredholm index of Toeplitz operators. The situation is,
however, quite different for Schr\"odinger operators with periodic
potentials where $PUP(E)$ failes to be Fredholm on intervals of
``energy bands" and where the Fredholm index in adjacent gaps can
jump by large integers \cite{tknn}.

\subsection{An open problem}

For applications to the Hall effect one considers $PUP$ (on the
range of $P$) where the projection $P$ depends on a parameter such
as the Fermi energy or the external magnetic field. The family
$PUP$ is therefore defined on different spaces, since the range of
$P$ is not fixed. Our strategy, so far, has been to study a family
of operators on a {\em fixed} Hilbert space. To adapt the QHE to
this strategy one must replace $PUP$ by something like
\be
C=PUP+1-P, \label{C}
\ee
acting on the full Hilbert space, as
$Index(C)$ on the full space coincides with
$Index(PUP)$ on $Range(P)$. Now, a deformation of $P$ leads to a
deformation of $C$ and gives a family of bounded operators on a
fixed space, say, $L^2(\complex)$. However, this modification is
not without a price since now, even for the simple case of a full
Landau level, $C$ is not strictly a Toeplitz operator. It is a
rather silly generalization of a Toeplitz operator to a direct sum
of a Toeplitz operator and the identity.

A more serious problem has to do with what should one pick as a
good family $P$. In particular,  when one considers a variation of
the Fermi energy $E$ the corresponding projection $P(E)$ is not
continuous in the operator norm. Hence, a smooth variation of $E$
is not even a smooth variation of $C$ in the operator norm (much less
in the sharper norms considered above).

Using the fact that the Fredholm index does not change under small
changes in the norm of the operator, there is no harm done if one
replaces the spectral projection $P(E)$ by the Fermi function
\be
P_\beta(E)= {1\over \exp \beta (H-E)+1}, \ee for $\beta$ large.
Unlike $P(E)$, $P_\beta(E)$ is a smooth function of $E$, and
so the family $C_\beta(E)$ is smooth. The price one
pays is that $P_\beta(E)$ is not a projection, which leads
to ambiguities as to what one might want to choose for
$C_\beta(E)$ . For example, instead of (\ref{C}) one might choose
\be
C_\beta(E)=P_\beta(E)UP_\beta(E)+(1-P_\beta^2(E)).
\ee
The trouble
is that it is not clear what, if anything, the results about
families of Toeplitz operators imply for the family $C_\beta(E)$.

We therefore pose the following questions:

For random Scr\"odinegr operators
on the plane, with $\beta$ sufficiently large, what are the properties of the
family of operators $C_\beta(E)$?  Is it Fredholm away from a discrete set of
energies $E$, or does it fail to be Fredholm on bigger sets?
If it fails to be Fredholm at isolated points, are the jumps
generically small?

\section*{Acknowledgments}
 This research
was supported in part by the Israel Science Foundation, the Fund
for Promotion of Research at the Technion, the DFG, the
National Science Foundation and the Texas Advanced
Research Program.

\end{document}